\documentclass[12pt]{article}
\usepackage{amsmath,amssymb,epsfig}
\RequirePackage[english,russian]{babel}
\RequirePackage[cp1251]{inputenc}

\begin{document}

\def\refname{References}

\begin{center}
{\Large
\textbf{Microscopic Approach in Inelastic Heavy-Ions Scattering
with Excitation of Nuclear Collective States}
}   %
\end{center}

\begin{center}
\textbf{K.V. Lukyanov${}^{1}$, E.V. Zemlyanaya ${}^1$, I.N.
Kuchtina${}^1$, V.K. Lukyanov${}^1$,\\
Z. Metawei$ {}^2$, K.M. Hanna${}^{1,3}$}
\end{center}

\noindent {\it 1 Joint Institute for Nuclear Research, Dubna, Russia \\%
2 Physics department, Faculty of Science, Cairo University, Giza, Egypt \\%
3 Math. and Theor.Phys. Dept., NRC, Atomic Energy Authority,
Cairo,Egypt \\%
} %

\vspace{0.2cm}

In the density distribution of a deformed target-nucleus, the
spherical $\lambda = 0$ and the deformed $\lambda = 2$ parts were
considered. On this basis, the corresponding potential parts $U_0$
and $U^{(2)}_{int}$ of a double -folding microscopic
nucleus-nucleus optical potential are obtained. Then, for these
potentials and by using the coupled-channel technique (ECIS), the
elastic and inelastic amplitudes are calculated for ${}^{17}O$
heavy ion scattering on $2^{+}$ collective excited state of
various target nuclei. Besides, the same cross-sections are
calculated in the frame of an adiabatic approach of the eikonal
approximation, where the inelastic amplitude is the linear
function of $U^{(2)}_{int}$. Both the obtained results are
compared with the experimental data, and also discuss their
efficiency in predicting the deformation parameters of nuclei.

\vspace{0.2cm}


\section{Introduction}

In general, it is well known the standard approaches for the
calculation of the inelastic scattering of nucleons from nuclei
accompanied with excitation of collective states [1]. For this
purpose, one needs a transition interacting potential $U_{int}(r)$
which includes both the Coulomb and nuclear parts. The Coulomb
part is obtained by using the standard method of multi-pole
expansion, while the nuclear part can be gained by an expansion of
an optical potential within the deformation addition to the
nuclear radius $\delta R({\hat r})$ such that to verify the
relation, $U(R+\delta R, r)=U(R,r)+U_{int}({\bf r})$. At small
deformations we have,$U_{int}=(dU(r)/dR)\delta R({\hat r})$.
Further, the calculation of the inelastic scattering amplitude can
be fulfilled either in the linear approximation of $U_{int}$, or
by using the coupled channel method where all orders of $U_{int}$
are included. In the investigation of nucleon-nucleus scattering,
it is natural to consider that the radius parameter $R$ and the
deformation are the same for the target nucleus and for the
scattering potential. Therefore, the deformation parameter of the
potential obtained by fitting with the experimental data is
concluded to be the deformation of the nucleus itself. In
principle, same prescriptions are used in calculation of inelastic
scattering of heavy ions. However, in so doing one should take
into account some specific considerations of such task. First, in
case of nucleus-nucleus scattering the deformation parameter of
the potential, generally speaking, does not coincide with that of
any single nucleus of these nuclei. Second, due to the charge
multiplication of nuclei $Z_1 Z_2 e^2$, the Coulomb field becomes
rather strong which demands to include the high orders of the
transition Coulomb potential. Third, with the increase of the
collision energy the usual standard methods suffer from
difficulties, due to the huge number of partial waves, which will
affect the preservation of the necessary accuracy. These
difficulties can be overcome by using the eikonal method of the
high energy approximation (HEA), which have been applied by us
before on analysis of both elastic [2,3] and inelastic [4,5]
scattering of nuclei and proved itself as an effective method. The
nuclear potential, as said above, was constructed in the form of a
derivative of some function either phenomenological as Woods-Saxon
type [4], or in the form of a microscopical folding potential of
elastic scattering [5]. Thus, the received value for the
deformation parameter after the comparison with experimental data
keeps the character of the nucleus-nucleus potential itself. In
Sec. 2 we construct the microscopical transition-density of one
the target nucleus, and then the corresponding transition
potential, and thus the nucleus deformation is  used as a fitting
parameter when compared with experimental data. Further, in Sec. 3
on the basis of this transition potential we calculate the
nucleus-nucleus inelastic cross section using the code ECIS for
numerical solution of the coupled channel equations as well as the
adiabatic approach in the HEA model which deals with an analytic
expression for inelastic scattering amplitude. In Sec. 4 a
discussion is carried out of results and the possibility to use
the heavy ion beams for obtaining information on the structure of
colliding nuclei.

\section{Folding Model of Transition Potential}

When constructing the transition microscopic potential we study
the excitation of the target nucleus "2". It is considered that
the density distribution function $\rho_2({\bf r}_2)$ includes
dependence on the collective coordinates of the nucleus
$\alpha_{\lambda\mu}$, which characterize coherent deformation of
the radius ${\bf r}_2$. In the case of quadrupole deformation
($\lambda=2$), the deviation from its spherical form $\rho_2(r_2)$
can be described by changing the spherical coordinate ${r}_2$ by
the coordinate ${\bf r}_2$ on the surface of an ellipsoid [1],
\begin{equation}\label{2.1}
r_2\Rightarrow {\bf r}_2 = r_2 + \delta {\bf r}_2,
\end{equation}
\begin{equation}\label{2.2}
\delta {\bf r}_2 = - r_2\sum_{\mu=0,\pm 2}\alpha_{2\lambda}\,
Y_{2\lambda}({\hat r}_2), \qquad \alpha_{2\lambda}=\beta
D^{(2)*}_{\mu 0}(\Theta_i),
\end{equation}
where ${\hat r}_2$ is the angular coordinates of ${\bf r}_2$ in
c.m system of colliding nuclei. Here, we consider the rotation
model of the nucleus and use as collective coordinates the Euler
rotation angles $\Theta_i$, while $\beta$ is the corresponding
deformation parameters. At $\beta\ll 1$ we keep only the linear
terms in the density expansion in $\delta {\bf r}_2$, so that
\begin{equation}\label{2.3}
\rho_2({\bf r}_2)\Rightarrow \rho_2(r_2) + o_2({\bf r}_2),
\end{equation}
\begin{equation}\label{2.4}
o_2({\bf r}_2) = o_2^{(2)}(r_2)\sum_{\mu=0,\pm 2}\alpha_{2\mu}
Y_{2\mu}({\hat r}_2), \qquad o_2^{(2)}(r_2)=-r_2
{d\rho_2(r_2)\over dr_2},
\end{equation}
where $o_2^{(2)}(r_2)$ defines the transition density form
\footnote{~From a series of papers one applies the model with a
deformation term $\delta {\bf r}_2 = -R_2\sum
\alpha_{2\lambda}Y_{2\lambda}({\hat r}_2)$, which leads to
$o_2^{(2)}(r_2)=-R_2 d\rho_2(r_2)/dr_2$. Here R$_2$ is considered
as the radius of the nucleus, although there is no accurate
definition for it. Due to this choice $o_2^{(2)}(r_2)$ usually
does not fit the deformation parameter, but the "deformation
length" $\delta_2=\beta R_2$ without factorizing separate
multipliers.}. Inserting the density given in (3) in the formulae
for direct and exchange parts of the folding potential (see e.g.
[3],[5]), the first term $\rho_2(r_2)$ leads to an expression for
the central part of the potential, while the second one $o_2({\bf
r}_2)$ expresses the quadrupole transition potential as follows,
\footnote {~ In the folding potential one usually introduces the
factor $F(\rho_1+\rho_2)$ to correct dependence of nucleon-nucleon
forces on the density of nuclei in their overlap region. We also
consider this factor in the central part of the potential, but do
not include it in the transition potential (5) because of the not
so clear of its physical interpretation in the case of inelastic
scattering.}
\begin{equation}\label{2.5}
V_{int}({\bf r})\,=\,V^D_{int}({\bf r})\,+\,V^{EX}_{int}({\bf r}),
\end{equation}
where,
\begin{equation}\label{2.6}
V^D_{int}({\bf r})\,=\,Cg(E)\,\sum_{\mu=0,\pm
2}\alpha_{2\mu}\,\int
\rho_1(r_1)o^{(2)}_2(r_2)\,v_{00}^D(s)\,Y_{2\mu}({\hat r}_2)\,d^3
r_1 d^3 r_2,
\end{equation}
\begin{equation}\label{2.7}
V^{EX}_{int}({\bf r})=Cg(E)4\pi \sum_{\mu=0,\pm 2}\alpha_{2\mu}
\int_0^\infty G^{(2)}({\bf r},{\bf
s})v_{00}^{EX}(s)\,j_0(K(r)s/M)\,s^2ds,
\end{equation}
\begin{equation}\label{2.8}
G^{(2)}({\bf r},{\bf s})=\int \rho_1(|{\bf u}-{\bf r}|) {\hat
j}_1(k_{F_1}(|{\bf u}-{\bf r}|)\cdot s)o^{(2)}_2(u)Y_{2\mu}({\hat
u})\, {\hat j}_1(k_{F_2}(u)\cdot s)d^3u.
\end{equation}
Here $j_n(x)$ is the spherical Bessel function, the function
${\hat j}_1(x)=(3/x^3)(\sin x-x\cos x)$, а $k_{F}$ is the Fermi
momentum of a nucleon in the nucleus, and $K(r)=[(2mM/\hbar^2)
(E_{cm}-V(r)-V_C(r))]$ is the local momentum of relative motion
which depends on the central part of the elastic scattering
potential calculated separately. In momentum representation the
integrals (6), (8) transform to integrals of one dimensional form
(see e.g. [6]), and the final expression for the transition
density can be given in the form
\begin{equation}\label{2.9}
V_{int}({\bf r})\,=\,V^{(2)}_{int}(r)\sum_{\mu=0,\pm
2}\alpha_{2\mu}\, Y_{2\mu}({\hat r}),
\end{equation}
\begin{equation}\label{2.10}
V^{(2)}_{int}(r)\,=\,V^{D(2)}_{int}(r)\,+\,V^{EX(2)}_{int}(r),
\end{equation}
where
\begin{equation}\label{2.11}
V^{D(2)}_{int}(r)\,=\,Cg(E){1\over 2\pi^2} \, \int_0^\infty
\rho_1(q)o^{(2)}_2(q)\,v_{00}^D(q)\,j_2(qr)\,q^2dq,
\end{equation}
\begin{equation}\label{2.12}
V^{EX\,(2)}_{int}(r)\,=\,4\pi\,C\,g(E)\,\int_0^\infty
G^{(2)}(r,s)\,v_{00}^{EX}(s)\,j_0(K(r)s/M)\,s^2ds,
\end{equation}
\begin{equation}\label{2.13}
G^{(2)}(r,s)\,=\,\int_0^\infty
h_1(q,s)h^{(2)}_2(q,s)\,j_2(qr)\,q^2dq,
\end{equation}
\begin{equation}\label{2.14}
h_1(q,s)\,=\,4\pi\int_0^\infty\,\rho_1(r)\, {\hat
j}_1(k_{F,1}(r)\cdot s)j_0(qr)\,r^2dr,
\end{equation}
\begin{equation}\label{2.15}
h^{(2)}_2(q,s)\,=\,4\pi\,\int_0^\infty o^{(2)}_2(x)\, {\hat
j}_1\bigl(k_{F,2}(x)\cdot s\bigr)\,j_2(qx)\,x^2dx,
\end{equation}
The Fourier transform functions without/with upper index "2" are
defined as follows,
\begin{equation}\label{2.16}
\rho(q) = 4\pi \int_0^\infty f(r) j_0(qr) r^2dr, \quad
f^{(2)}_2(q) = 4\pi \int_0^\infty f^{(2)}_2(r) j_2(qr) r^2dr,
\end{equation}

\section{Inelastic Scattering Cross Sections Calculations}

On the basis of the above formulae we calculate microscopical
transition potentials and then use them to get the cross sections
of inelastic scattering for different nuclei. It is necessary that
the central part of the optical potential to be known which will
explain the experimental data of elastic scattering for the same
colliding nuclei at the same colliding energy in the inelastic
scattering channel. Calculations were performed in both cases,
first by using ECIS numerical code for coupled channels [7] as
well as in the adiabatic approach in HEA [4,5]. In HEA approach it
is supposed an adiabatic process when the velocity of an internal
nuclear motion is much smaller than the relative motion velocity
of the nucleus as a whole. Then the process can be considered as
if a scattering from "frozen" nucleus with fixed collective motion
coordinates $\{\alpha_{\lambda\mu}\}$. The use of HEA in this case
means the calculation of eikonal amplitude $f_{el}$ of elastic
scattering, where the potential has central and quadrupole parts
and depends on $\{\alpha_{\lambda\mu}\}$. The inelastic scattering
amplitude is constructed in the form of a transition matrix
element from initial $|00\rangle$ to final (excited) state of the
nucleus $\langle IM|$ where the operator is the elastic scattering
amplitude,
\begin{equation}\label{3.17}
f_{in}(q)\,=\,\langle IM| f_{el}(q,\{\alpha_{\lambda\mu}\})
|00\rangle
\end{equation}
Usually in this approach there is inherent approximation when the
part of the eikonal phase, $\chi_{int}=-(1/\hbar v)\int
V_{int}dz$, defined through the transition potential, is
considered of small value and then the corresponding part of the
eikonal $\exp(i\chi_{int})$ can be expanded in a series and limit
ourselves to the linear terms of $V_{int}$. Then using (9)the
amplitude (17) can be factorized to reveal a structure factor
$F_{\lambda\mu}= \langle IM|\alpha_{\lambda\mu}|00\rangle$. In
this case, for a given $\lambda$, the angular distribution does
not depend on the collective nature of the excited state, because
the structure factor affects only the absolute value of the cross
section. As is known from the nucleus rotational model, the
transition structure factor 0${^+}\to$2${^+}$ is equal to
$F^{rot}_{2,\mu}= \beta(1/\sqrt{5})\delta_{M,\mu}$ which reflects
that the absolute value of the cross section depends only on the
deformation parameter $\beta$. All the related details of formulae
can be find in [4,5]. The calculation procedure has two steps.
First, a microscopic calculation for both of the real $V$ and the
imaginary $W$ parts of the central optical potential is carried
out to compute the elastic scattering cross sections. This latter
is made with the help of code ECIS [7] and also using the HEA
model the differential cross sections, and then they are compared
with the experimental data. During the fitting process a variation
will be done for evaluating the strength parameters $N_r$ and
$N_{im}$ which define contributions of the real $V$ and the
imaginary $W$ parts into microscopic optical potentials. As a
result we denote optical potentials, in the elastic channel as
follows,
\begin{equation}\label{3.18}
U_{opt}(r)\,=\,N_rV(r)\,+\,iN_{im}W(r).
\end{equation}
The elastic scattering of $^{17}$O nucleus from $^{60}$Ni,
$^{90}$Zr, $^{120}$Sn and $^{208}$Pb nuclei at energy
E$_{lab}$=1435 MeV was studied previously by us [5] using
corresponding potentials in HEA case and results were compared
with experimental data given in Ref [8]. Similar calculations by
using ECIS code are given in [9]. The real parts of these
micro-potentials are calculated using the double folding model
which includes an exchange term with the effective nucleon-nucleon
Paris potential in the form of the CDM3Y6 dependence on density
[6](see also [5]). As to the imaginary part $W^H$, it was
calculated first in the frame of microscopic approach, given in
details in Ref [3], on the basis of the Glauber and Sitenko
multiple scattering theory [10,11]. In addition, in [3] it was
used for the imaginary part, as another version, the form of the
double folding real part $V^{DF}$. It was clear from all these
different types of calculations, that by fitting the strength
parameters $N_r$ and $N_{im}$ one can get the respective cross
sections in a good agreement with the experimental data. However,
in our present consideration we prefer to apply to the double
folding transition potential $V_{int}$ as the sample for the
imaginary potential when fitting the normalizing strength
parameters $N_r$ and $N_{im}$. So in the following we use the
microscopic transition potential in the form,
\begin{equation}\label{3.19}
U_{int}(r)\,=\,N_rV_{int}(r)\,+\,iN_{im}V_{int}(r),
\end{equation}
where $V_{int}$ potential is calculated by using Eqs.(9)-(15). In
Fig. 1 it is shown the radial dependence of the transition
potentials $N_rV^{(2)}_{int}$, and also the central potentials of
elastic scattering $N_rV^{DF}$. In concern of the transition
Coulomb potentials, they were calculated as usual for the
interaction of a charge $Z_1e$ with the field of uniformly
distributed charge $Z_2e$ in ellipsoid volume with radius
$R_c\left(1+\sum \alpha_{2\mu} Y_{2\mu}({\hat r})\right)$, where
$R_c=1.2\left(A_1^{1/3}+A_2^{1/3}\right)$ fm. The corresponding
relations for central and quadrupole transition Coulomb potentials
and also for the eikonal phases are given in [4].

\begin{figure}
\begin{center}
\psfig{file=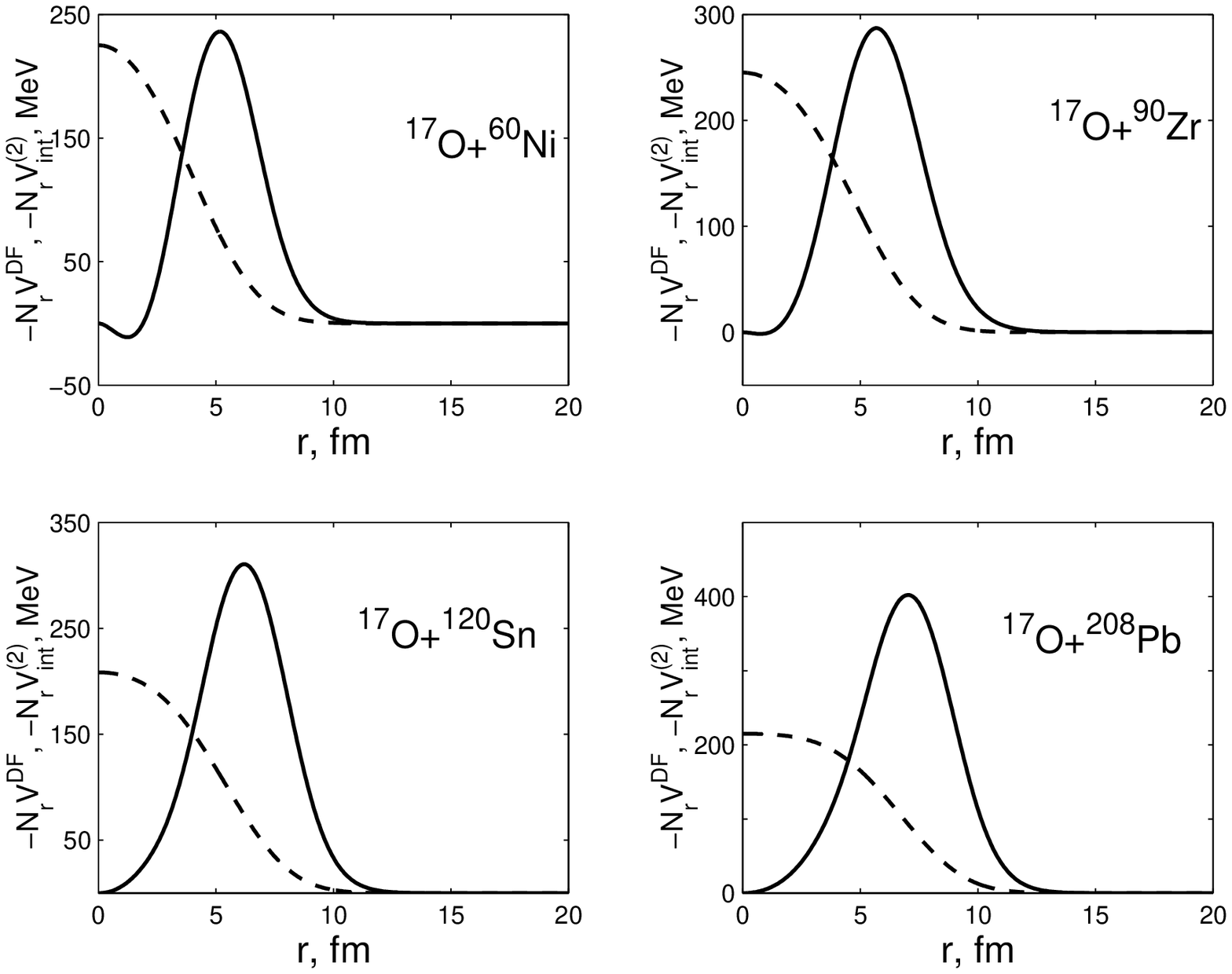,width=.95\linewidth}
\end{center}
Fig. 1. Real part of potentials $N_rV^{DF}(r)$ of elastic (dashed
lines) and $N_rV^{(2)}_{int}(r)$ of inelastic (solid lines)
nucleus-nucleus interaction, for the scattering of heavy ions
${^{17}O}$ from different target nuclei at $E_l=1435$ MEV( see
text).
\end{figure}

\begin{table} {\large {\bf Table.}
Renormalizing Parameters $N_r,N_{im}$, Coulomb Deformation
$\beta^{(c)}$ and $\beta^{(n)}$, and Nuclear Transition Potentials
Parameters} \\[.3cm]
\begin{tabular}{|c|c|c|c|c|c|}
\hline \hline

Parameters        &      &$^{17}$O+$^{60}$Ni &$^{17}$O+$^{90}$Zr
                       &$^{17}$O+$^{120}$Sn &$^{17}$O+$^{208}$Pb \\
\hline \hline
Renormalizing  &$N_r$& 0.6 & 0.6  & 0.5 & 0.5 \\
\cline{2-6}

coefficients   &$N_{im}$& 0.6 & 0.5 & 0.5 & 0.8  \\
\hline \hline
Coulomb deformation    &$\beta^{(c)}$& 0.2067 & 0.091  & 0.1075 & 0.0544 \\
\hline
 nuclear deformation (ECIS)       &$\beta^{(n)}$&  0.2541 &  0.071 & 0.1063   & 0.078$^{\ast}$     \\
\hline
 nuclear deformation (HEA)       &$\beta^{(n)}$&  0.4   &  0.14  & 0.25   & 0.12
 \\                       
\hline \hline \end{tabular} \\ [.3cm]

$^{\ast}$~For the imaginary part value of the transition density
$\beta^{(n)}_{im}=0.0222 $ \\ [1cm]
\end{table}

\begin{figure}
\begin{center}
\psfig{file=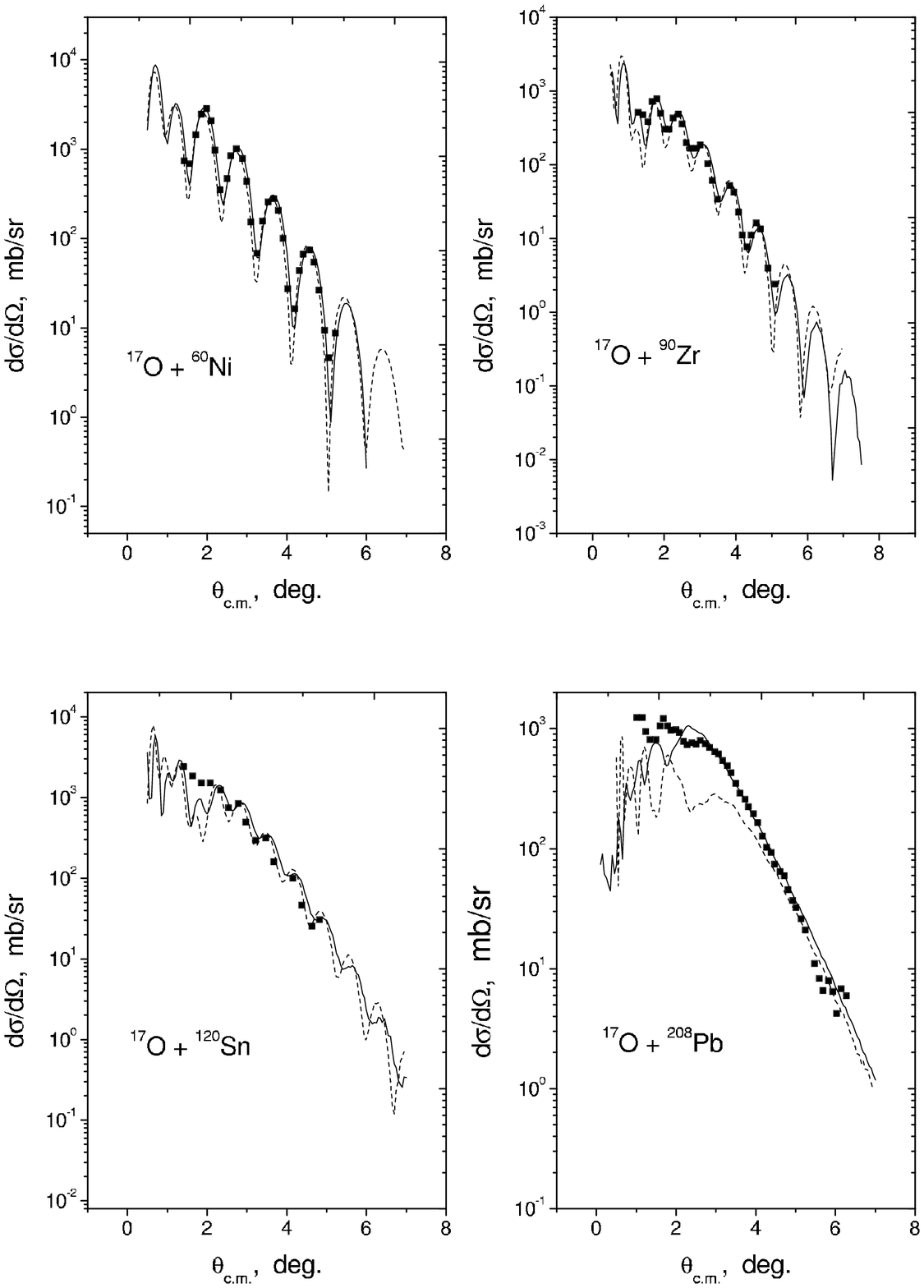,width=.8\linewidth}
\end{center}
Fig. 2. Inelastic differential cross section, calculated by using
coupled channels method (solid lines) and HEA method(dashed lines)
for interaction of nucleus ${^{17}O}$ with target nuclei at 1435
MeV with excitation of transitions $0^+\to 2^+$. Experimental data
from [8].
\end{figure}

\begin{figure}[t]
\begin{center}
\psfig{file=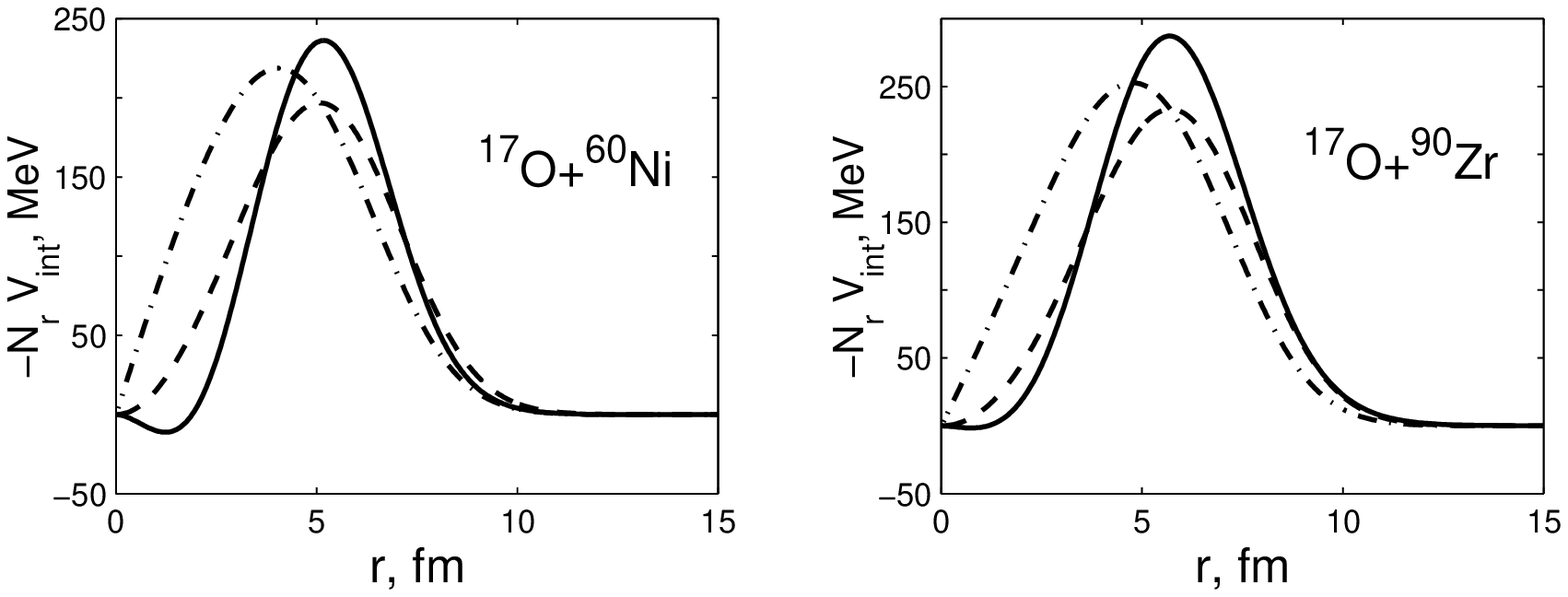,width=.95\linewidth}
\end{center}
Fig. 3. Transition potentials (radial part) for inelastic
scattering of heavy ions ${^{17}O}$ from nuclei ${^{60}}$Ni and
${^{90}}$Zr at 1435 MeV. Microscopic calculations
$N_rV^{(2)}_{int}(r)$ -- solid lines. Semi-microscopic model :
$V^{DF-I}_{int}(r)$ (dashed) and $V^{DF-II}_{int}(r)$ (dash
dotted) (see text).
\end{figure}

\begin{figure}[t]
\begin{center}
\psfig{file=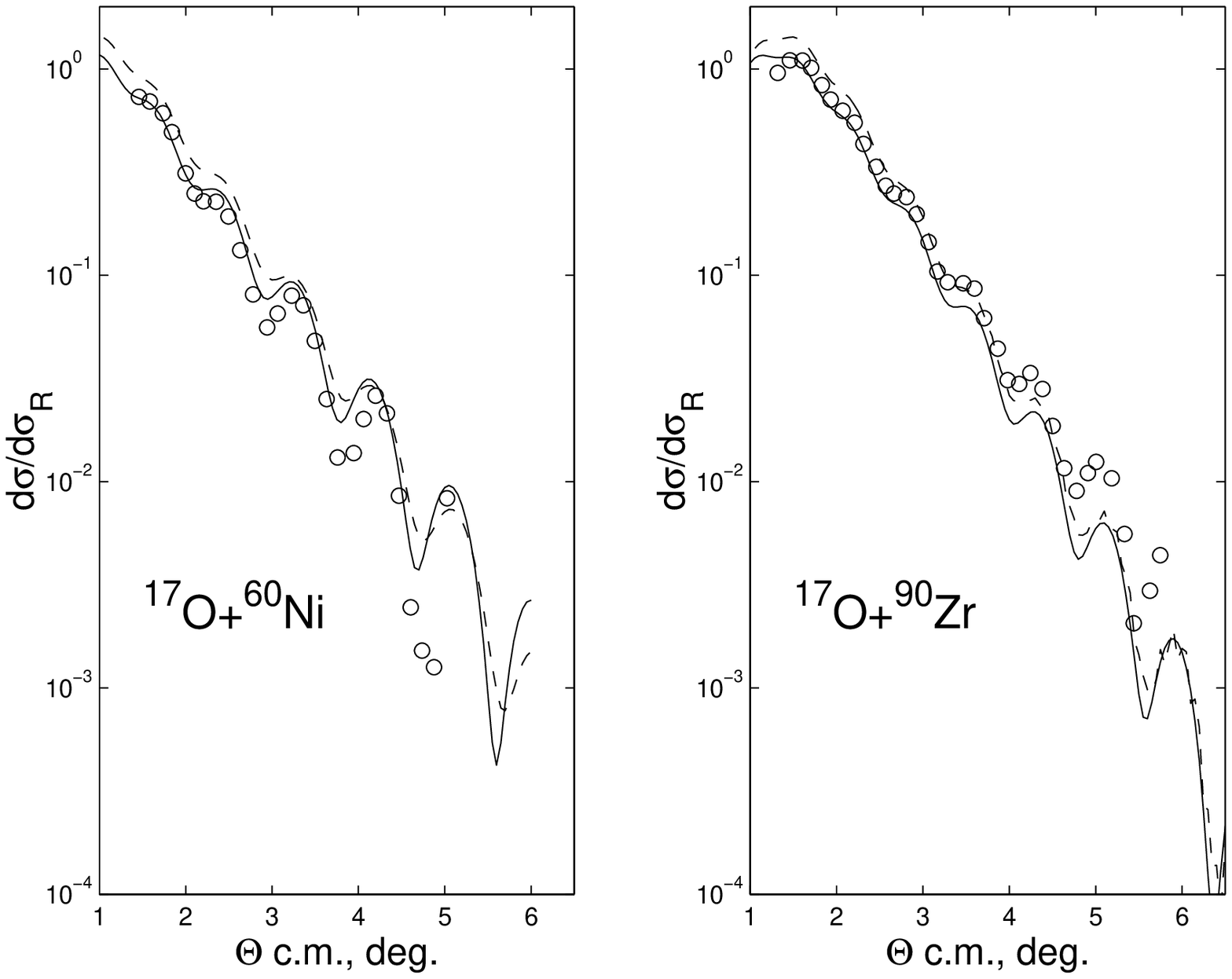,width=.95\linewidth,height=10cm}
\end{center}
Fig. 4. The ratio of elastic differential cross sections to the
Rutherford ones for scattering of heavy ions ${^{17}}$O on nuclei
${^{60}}$Ni and ${^{90}}$Zr at 1435 MeV without coupled channel
effects $0^+\to 2^+$ (solid lines) and with this effect (dashed
lines). Experimental data from [8].
\end{figure}

\section{Comparison with Experimental Data, Discussion,
and Conclusions}

In Fig. 2 it is shown the theoretical calculations for the
differential cross sections of inelastic scattering using the code
ECIS (solid lines) and HEA (dotted lines). The elastic scattering
potential therewith is calculated, as said above, by fitting the
strength parameters $N_r$ and $N_{im}$ to the experimental data
without coupling between the elastic and inelastic channels. Next
step is calculations of inelastic scattering by fitting only the
deformation parameter $\beta$ for the target nucleus. It is to be
noted that when using the ECIS code for inelastic scattering the
coupling of elastic and inelastic channels was involved through
the transition potential, and its contribution was taken in all
orders of perturbation theory. This contribution is defined by the
deformation parameter in the density distribution function
$\beta^{(n)}$ of the target nucleus and by the deformation
$\beta^{(c)}$ of the unified charge density distribution in a
sphere with radius as the sum of radii of colliding nuclei. In
general, this parameter must not coincide with the deformation
parameter of the charge distribution of the target nucleus.
Nevertheless, we get the value of $\beta^{(c)}$ from the data of
electric transition probabilities to $2^+$ level for different
target nuclei as given in Ref.[8]. These values are done in the
relevant Table. The second parameter $\beta^{(n)}$ was fitted to
the experimental data of inelastic scattering. As to the HEA
calculations of cross sections, these parameters are involved as
the linear terms of the transition potentials in the inelastic
scattering amplitude. We notice from Fig. 2 that in both cases one
gets reasonable good agreement with experimental data, except at
the small scattering angles in the HEA calculations. The received
results for the deformation parameters $\beta^{(n)}$ in the two
methods differ in 1.5 - 2 times from each others. In all cases
except the scattering on $^{208}$Pb nucleus in ECIS calculations,
the $\beta^{(n)}$ parameters for real and imaginary parts of the
transition potentials are equal. As an exceptional case we were
obliged to perform fit with different deformation parameters for
both of the real and imaginary parts of the transition potentials
and we received the following values $\beta_r^{(n)}$=0.078 and
$\beta_{im}^{(n)}$=0.0222. However, it is conceivable that this
result can be changed in case if one changes the transition
density $o^{(2)}_2$ from the form
$r_2\left(d\rho_2(r_2)/dr_2\right)$ to the form
$R_2\left(d\rho_2(r_2)/dr_2\right)$, where $R_2$ is the radius of
the target nucleus. In fact, this leads to enlarge the
contribution of the Coulomb transition potential in the peripheral
region, and therefore one does not enhances the nuclear part of
the amplitude at the sacrifice of the imaginary part of the
transition nuclear potential. This is shown in Fig. 3 where it is
given three transition potentials, two of them (dashed and dotted
dashed) correspond to the transition potentials constructed from
derivatives of the microscopic potentials of elastic scattering in
the forms $V^{DF-I}_{int}=N_r\,r\left(dV^{DF}(r)/dr\right)$ and
$V^{DF-II}_{int}=N_r\,R\left(dV^{DF}(r)/dr\right)$, where $R=r_m$
is the radius at the maximum point of the function
$V^{DF-I}_{int}$. In the peripheral area, the second transition
potential occurs lower in comparison with the first one, giving
the advantage to the Coulomb transition potential content, which
illustrates the above mentioned comments. The third solid curve is
our microscopic potential $N_rV^{(2)}_{int}$ and we see the
difference in its behavior in comparison with the other two
potentials constructed on the basis of semi-microscopic approach
where the transition potentials are constructed as the derivatives
of elastic scattering potentials.

In conclusion, we denote that it was better to define
simultaneously all the three parameters $N_r$, $N_{im}$, and
$\beta^{(n)}$ in the process of fitting for getting the cross
sections of elastic and inelastic scattering, in place of dividing
the problem into two steps, the first to get the strength
parameters $N_r$ and $N_{im}$ in the elastic channel and then the
$\beta^{(n)}$ parameter in the inelastic channel. As an example,
in Fig. 4 the solid lines show the differential cross sections of
elastic scattering when the micro-potentials the $N_r$ and
$N_{im}$ parameters are from the Table while the dashed line shows
the cross section when we use the same potentials and
simultaneously takes into account the effect on elastic scattering
of the inelastic channel in the ECIS code. The effect of inelastic
channel does not give, in principle, an observable change of the
angular distribution in elastic cross sections. In view of the
above study it can be seen, in general, the ways of the further
study of mechanisms of nucleus-nucleus inelastic scattering to
receive more accurate data about the deformation parameters and
also on the transition structure matrix elements.

The co-authors K.V.L, E.V.L and I.N.K are grateful to the Russian
Foundation for Basic Research (project 06-01-00228) for support.
Also deep gratitude from K.M. Hanna to both the authorities of
AEA-Egypt and JINR-Russia for their support.


\end{document}